\documentclass{vgtc}                          

\ifpdf
  \pdfoutput=1\relax                   
  \usepackage{graphicx}                
  \DeclareGraphicsExtensions{.pdf,.png,.jpg,.jpeg} 
\else
  \usepackage{graphicx}                
  \DeclareGraphicsExtensions{.eps}     
\fi%

\graphicspath{{figures/}{pictures/}{images/}{./}} 

\usepackage{microtype}                 
\PassOptionsToPackage{warn}{textcomp}  
\usepackage{textcomp}                  
\usepackage{mathptmx}                  
\usepackage{times}                     
\usepackage{cite}                      
\usepackage{tabu}                      
\usepackage{booktabs}                  
\usepackage{siunitx}

\onlineid{0}

\vgtccategory{Research}

\vgtcinsertpkg

\title{Line Harp: Importance-Driven Sonification for Dense Line Charts}

\author{Egil Bru\thanks{E-mail: egil.bru@student.uib.no} \\ 
\parbox{1.4in}{\scriptsize \centering University of Bergen, \\ Norway}
\and Thomas Trautner\thanks{E-mail: thomas.trautner@uib.no} \\
\parbox{1.4in}{\scriptsize \centering University of Bergen, \\ Norway}
\and Stefan Bruckner\thanks{E-mail: stefan.bruckner@uni-rostock.de} \\
\parbox{1.4in}{\scriptsize \centering University of Rostock, \\ Germany}
}

\teaser{
  \centering
  \includegraphics[width=\linewidth]{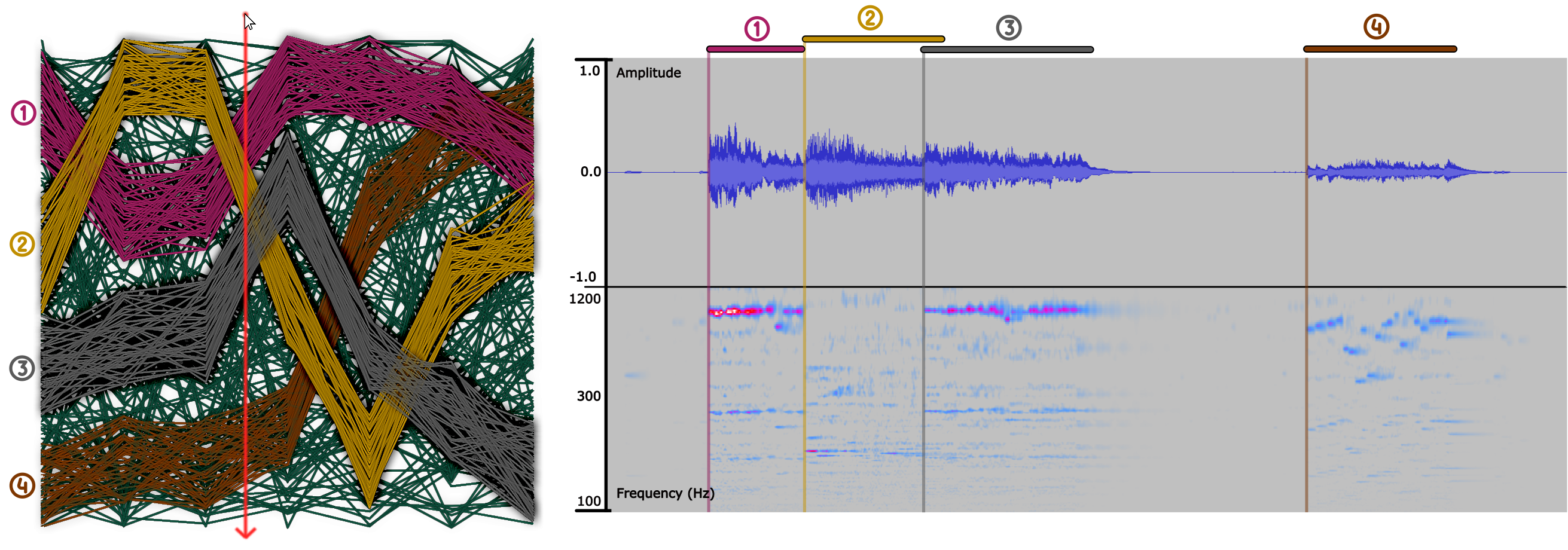}
  \caption{Dense line chart visualization containing multiple clusters, where a red arrow indicates the mouse path. On the right side, an audio visualization of our sonification approach is displayed for the highlighted mouse path. It contains an amplitude visualization and the contour
of the fundamental frequency (musical pitch) of the audio (see Section~\ref{sec:results} for further details).}
  \label{fig:teaser}
}

\abstract{Accessibility in visualization is an important yet challenging topic. Sonification, in particular, is a valuable yet underutilized technique that can enhance accessibility for people with low vision. However, the lower bandwidth of the auditory channel makes it difficult to fully convey dense visualizations. For this reason, interactivity is key in making full use of its potential. In this paper, we present a novel approach for the sonification of dense line charts. We utilize the metaphor of a string instrument, where individual line segments can be "plucked". We propose an importance-driven approach which encodes the directionality of line segments using frequency and dynamically scales amplitude for improved density perception. We discuss the potential of our approach based on a set of examples.
} 

\CCScatlist{
  \CCScatTwelve{Human-centered computing}{Visualization}{Visualization application domains}{Information visualization}; 
  \CCScatTwelve{Human-centered computing}{Human computer interaction (HCI)}{Interaction techniques}{Auditory feedback}; 
  \CCScatTwelve{Human-centered computing}{Accessibility}{Accessibility systems and tools}{}
}

\begin{document}


\firstsection{Introduction}

\maketitle

While there are several guidelines and best practices for designing accessible visualizations \cite{AccessibilityWild,AccessibleVisualization}, some users may still face barriers due to their individual needs and abilities. Sonification has the potential to enhance visualization accessibility by providing an additional channel to encode information. It has been employed in various real-world scenarios, including air traffic control \cite{AirTraffic} and medical equipment \cite{medical}. Real-time monitoring applications that use sonification have been extensively researched, including EEG signal monitoring \cite{EEG}, and stock market analysis \cite{Marketbuzz}. However, since auditory information allows for a limited degree of parallelism, it is difficult to convey complex information in a responsive and efficient manner. A static playback of dense data would either need to be excessively lengthy or too fast for proper perception, or, alternatively, require considerable abstraction.

In this paper, we present a new interactive technique for sonifying dense line charts. Our approach draws inspiration from string instruments like the harp, where each string produces a unique sound. Based on this analogy, we have designed a sonification strategy that mimics the "plucking" action when the mouse cursor intersects a line segment. To handle situations where multiple line segments are intersected simultaneously or in rapid succession, we have developed an importance-driven approach that adjusts the waveform based on the respective importance of line segments. This leads to the creation of "chords" in regions of dense data, enriching the auditory experience. Users can control the movement speed of the cursor, allowing them to either obtain a quick overview of the data set's larger-scale features or delve into the finer details of individual line segments.


\section{Related Work}


The Sonification Handbook \cite{TheSonificationHandbook} provides an extensive review of sonification theory, practice, and applications, featuring contributions from various researchers in the field.


Sonification intended for data analysis commonly involves a technique known as parameter mapping which involves the association of data elements with auditory parameters. Previous work has investigated the effectiveness of different mappings \cite{THIRTEENYO} and explored what types of data and polarities that are more naturally mapped to particular sound attributes \cite{SonificationMappings, SonificationMappings2}. Flowers and Hauer have demonstrated that participants can interpret line graphs through sonification, by mapping each data point to a musical note (frequency) and moving along the x-axis (time) \cite{VisualGraphs}. In one study, participants were able to successfully redraw multiple lines and find intersections while listening to two separate lines in each ear \cite{Drawing_by_Ear}. Furthermore, spatial audio has been proposed for visualizing multiple data series \cite{Design_Guidelines} in parallel. However, the effectiveness of such an approach is severely limited by the density of the data. As with visual clutter, audio noise increases as more lines are sonified, making it challenging for users to discern individual data points. In addition, the potential for interactions is limited by the fact that the sonification is presented sequentially. Common concepts such as pause, play, back, and forward may aid in the navigation of the graphs \cite{BrowsingModes}, but they do not fully address the limitations of the audio environment for complex visualizations.

To address this, model-based sonification approaches have been proposed \cite{TheoryOfSonification}. In model-based sonification, the data set is turned into a dynamic model to be explored interactively by the user, rather than sonifying the data directly. This approach to sonification was first proposed by Herman and Ritter \cite{Model-Based}. Based on interactive concepts from parameter mapping, Bovermann et al. \cite{TANGIBLESCANNING} suggested an interactive tool to change the sequence in which sonified data is presented. They propose the use of a tangible plane that can be oriented and positioned using touch interaction or sliders, allowing for the intersection of plane and data points. Intersection excites data objects enabling interactive auditory brushing. A similar approach employs the principal curve technique to compute a smooth path through the data set \cite{Hermann2000PrincipalCS}. This path can then be navigated by the user sequentially, providing audio feedback. Both techniques are heavily dependent on the density and distribution of the data.

Interactive lenses are a well-established class of visualization methods that facilitate multi-faceted data exploration \cite{Toolglass}. These lenses allow users to temporarily modify the visualization to display more details or different arrangements \cite{Interactive_Lenses}. In particular, lenses can be used to distort edges in graphs, reducing edge congestion, or to distort three-dimensional data, clearing a visual path to the focus \cite{EdgeLens}. Our approach combines established lens techniques with the power of sonification. Herman and Ritter \cite{Model-Based} proposed using an advancing shock wave to sonify data points that intersect with the wave front. This technique was later expanded upon with multi-touch interactions \cite{Tnnermann2009MultitouchIF}. This approach to sonification is particularly effective for lenses, as shown by Enge et al. \cite{SoniScope20221095}, where a shock wave was used within the lens to sonify intersecting data points in focus. However, these approaches are less suitable for line charts as they do not effectively convey relationship information.

\section{Line Harp}
While previous research has explored sonification of line charts using mostly static parameter mapping \cite{Drawing_by_Ear, VisualGraphs}, we believe that an interactive approach is more suitable for the characteristics of dense line charts. We base our approach on the importance-driven visualization technique for dense line charts by Trautner and Bruckner \cite{LineWeaver}. This method uses the notion of an importance function associated with each line, which allows them to interweave individual lines such that the most important lines segments occlude those with lower importance. Similarly, we use importance to control the amplitude such that more important line segments will be played louder than those with a lower importance value. The frequency of sound generated by plucking a string in a real-world musical instrument is influenced by various factors, such as the material properties of the string and the tension it is under. For the purpose of sonification, we use frequency to encode the directionality of the corresponding line segment. In line charts, with their common left-to-right reading order, this is relevant in the identification of salient features such as trends (e.g., an upwards or downwards development in a time series).

\subsection{Sonification}

Following Trautner and Bruckner \cite{LineWeaver}, we regard line data as set $D = \{L_1, L_2, ... , L_N\}$ of $N$ polylines with its members $L_i = (P_1, P_2, ... , P_M)$ represented as tuples of $M$ ordered two-dimensional points $P_i = (x_i, y_i)$. The resulting parametric curve $l_i(u)$ of each member is a polyline generated by linear interpolation between its associated points. Furthermore, we use their notion of an importance function $\beta_i(u) \in [0,1]$ which associates a scalar importance value with every position along the curve. In practice, there are many different ways to define importance, such as using underlying data, results of a features detection algorithm, or fundamental properties of the lines. In their paper, for instance, Trautner and Bruckner \cite{LineWeaver} present an algorithm that generates importance values based on a heuristic optimization of screen space utilization for multiple sets of lines within the same chart, and then use this information for importance-driven blending~\cite{Bruckner-2010-HVC}.

\begin{figure}[t]
 \centering 
 \includegraphics[width=\columnwidth]{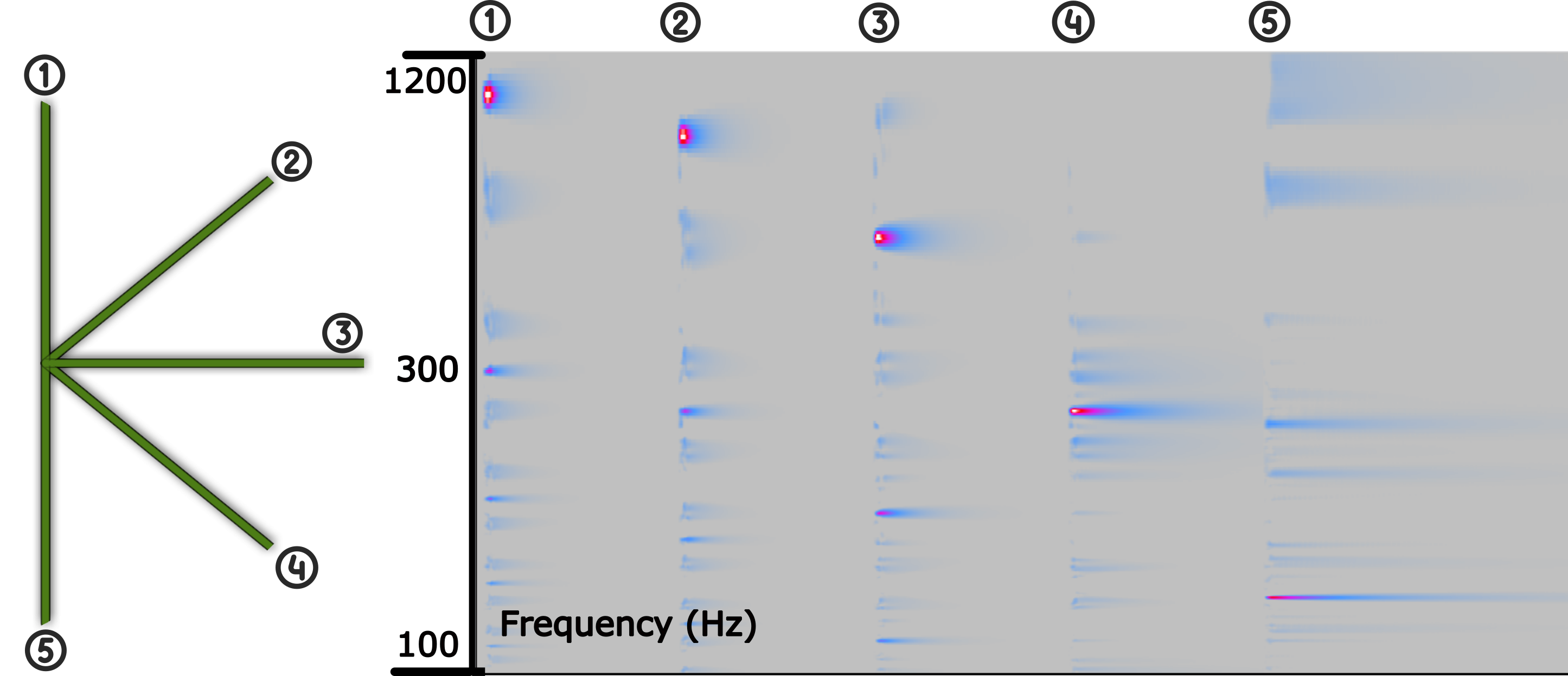}
 \caption{Visualized audio frequencies for lines 1--5 originating at the same point. The audio visualization is the contour of the fundamental frequency (musical pitch) of the audio, using the Enhanced Autocorrelation (EAC) algorithm as implemented in the spectrogram view of the popular audio software Audacity~\cite{audacity} (see Section~\ref{sec:results} for a more detailed description of this visualization).}
 \label{fig:angleMapping}
\end{figure}

"Plucking" such a line at any point generates a note defined by its frequency, amplitude and decay, where decay describes how the sound changes from peak amplitude to zero. As previously mentioned, we use the direction of a line segment to control the frequency of the generated note. In practice, frequencies are mapped to angles as illustrated in Figure~\ref{fig:angleMapping}. As common in line charts, we assume a left-to-right reading order, and hence straight up and down lines are mapped to the highest and lowest frequencies, respectively. In visually dense line charts, and especially parallel coordinates \cite{STAR_PC}, overplotting and overcrowding is a major issue. The issue also applies to sonified dense line charts. We address this by using the importance value to control the amplitude of the note, i.e., lines that are visually occluded due to their lower importance, will also be deprioritized in the sonification.

In most cases, auditory feedback is presented sequentially, which essentially requires the user to wait for the next note. However, this approach is not ideal as it contradicts the swift response time of visual interactions, which typically require a response within 50--100$\mskip\thinmuskip$ms \cite{Tominski20IVDA}. In musical terms this means playing notes within 1200--600$\mskip\thinmuskip$bpm (beats per minute). A study on auditory inspection time (IT), where participants where tasked with discerning the difference between a high tone and a low tone, found that 95 percent responded correctly when the frequency changed every 100--200$\mskip\thinmuskip$ms~\cite{ClockingTheMind}. Although this study focused on discerning two easily distinguishable frequencies, its findings suggest that slower changes in frequencies may not always be optimal for auditory feedback in interactive contexts like ours.

\begin{figure}[b]
 \centering 
 \includegraphics[width=\columnwidth]{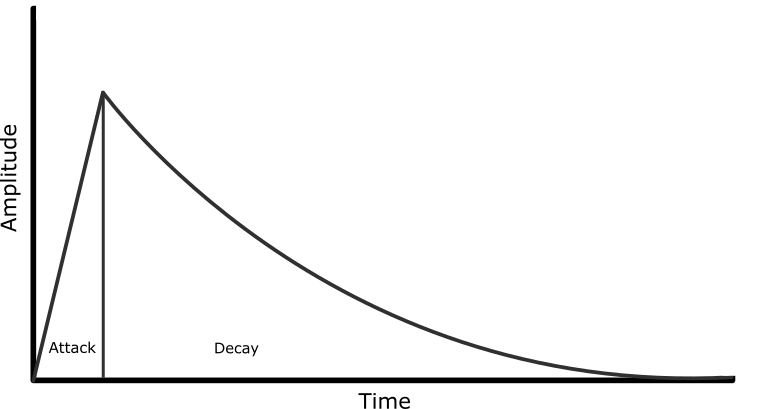}
 \caption{Attack and decay (AD) envelope used to produce a plucked string sound. Attack refers to the initial sound build up, while decay describes how the amplitude falls off over time.}
 \label{fig:envelope}
\end{figure}

We address this issue by the fact that most periodically vibrating objects to which we attribute frequency, including the human vocal folds and the strings of musical instruments, vibrate at several sinusoidal component frequencies simultaneously called the harmonic spectrum. Although each of these frequencies sounded alone has unique frequencies, when sounded simultaneously they perceptually fuse and collectively evoke a singular periodic frequency \cite{Music_Perception_2002}. This concept also applies to our analogy of a harp, where producing musical chords requires the plucking of multiple strings in combination. Practically speaking, this involves accumulating the sinusoidal components of all currently active notes, with "active notes" referring to all objects that are currently vibrating and producing sound. Objects that can produce sound, such as harp strings, have a more complex spectrum compared to pure sine waves. This complexity makes them more pleasant to listen to and easier to perceive \cite{Ramloll2001UsingNS}. In our approach we therefore use synthesized instruments to produce sound.

In music and sound, the envelope describes how a sound changes over time, and different instruments may have varying envelopes (see Figure \ref{fig:envelope}). For example, a piano key, when struck and held, creates a near-immediate initial sound which gradually decreases in volume to zero. To convey the acoustic impression of our lines, we opted to use the sound of a plucked string instrument. In previous studies, researchers have taken various approaches, such as using different instruments to represent distinct lines \cite{Drawing_by_Ear}, or assigning different instruments to separate clusters \cite{SONIFICATION_DENSEDATA}. In contrast, our approach involves using a single instrument that maintains consistency throughout, with only the frequency and amplitude varying. This approach was taken due to the fact that not all instruments are capable of producing all frequencies, and even when they can, the resulting sound may vary based on the frequency \cite{guidelinesForEarcons}. Furthermore, the plucked string also conveys a clear pitch and has a fitting decay for our purposes.

\begin{figure}[t]
 \centering 
 \includegraphics[width=\columnwidth]{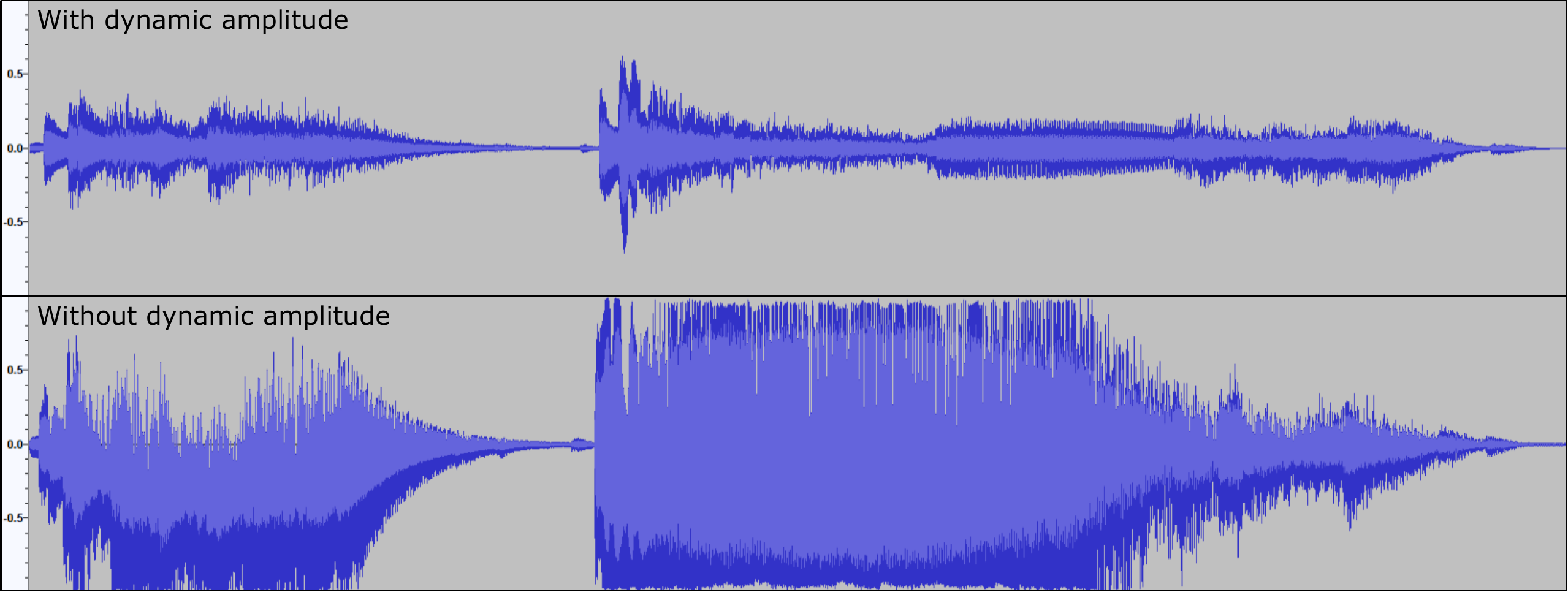}
 \caption{Top: audio with dynamic amplitude scaling. Bottom: audio without any amplitude and decay scaling. The duration and notes are the same for both cases. The gaps in the top audio are due to clipping. It can be seen that the clusters are not discernible without dynamic amplitude scaling.}
 \label{fig:comparison}
\end{figure}

\begin{figure}[b]
 \centering 
 \includegraphics[width=\columnwidth]{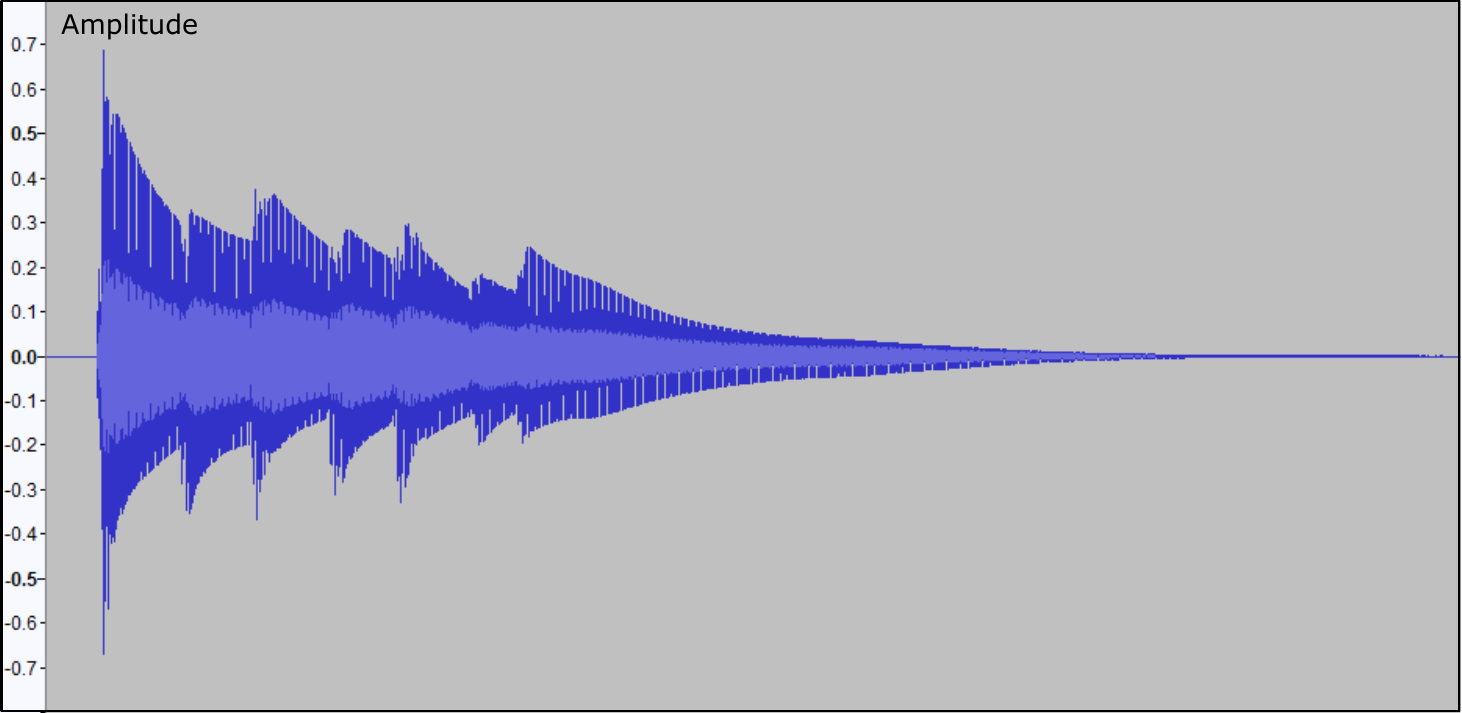}
 \caption{Audio feedback produced from evenly spaced lines played in succession demonstrating how the initial note is accented and the last note is played slightly longer.}
 \label{fig:combined}
\end{figure}

In the context of a plucked string instrument, the decay, or duration, refers to the amount of time it takes for the amplitude to decrease from its peak to zero. An AD (Attack, Decay) envelope is commonly used for these types of instruments, with attack referring to the time taken from key press to peak amplitude. For our purposes, we want a near-immediate initial sound (Attack) and a relatively quick initial decay that gradually fades away to zero.  This envelope is illustrated in Figure \ref{fig:envelope}. However, a simple accumulation of the produced sound would lead to audio clipping and distorted output, as seen in the lower part of Figure~\ref{fig:comparison}, as our approach does not wait for each note to finish. To address this issue, we employ a dynamically scaled amplitude and decay. Specifically, the amplitude is scaled based on the total cumulative amplitude of all active notes, while the decay is proportionally reduced based on the number of active notes being played. In practice, this means that when playing loud (important) lines quickly, they do not become overly loud, while only playing  lines that are less important still results in a relatively loud sound. This ensures a balanced audio output that still accurately reflects the relative importance and density of the data. The dynamic decay ensures that transitions between different volume and frequency levels occur more quickly when more notes are played, making the output more responsive to changes caused by interaction. As a result, the audio feedback produced by the system accents the first note slightly louder and makes the last note slightly longer \cite{guidelinesForEarcons}. This concept is further demonstrated in Figure \ref{fig:combined}.


\subsection{Interaction}

Per default, whenever the mouse cursor is moved, we "pluck" all line segments that intersect the current location and generate notes according to the importance values at the point of intersection. The currently played lines are visually indicated using a vibration effect, to reinforce the musical instrument analogy, and can optionally be highlighted using color. 

We also offer several additional interaction facilities controlled by different key bindings. For example, in order to investigate regions where multiple clusters intersect, we offer an additional lens mode. The lens allows users to reveal lines with lower importance values. Visually, we employ a variation of the distortion lens approach by Carpendale et al.~\cite{Distortion3D} to displace lines with importance values above a certain threshold, while sonically, this amounts to only playing notes for lines below the threshold. Furthermore, we also provide a lens playback feature, which sonifies every line in the lens in importance order. This provides a quick and easy way to gain an overview of all lines within the lens radius.

\section{Implementation}

Our approach was implemented in C++ and OpenGL. The sonification uses the Gamma library (\url{https://github.com/LancePutnam/Gamma}), which is a generic synthesis library for C++. Gamma handles audio in a separate thread from the visuals, allowing for seamless integration. The library contains a \verb|Pluck| class, which produces a plucked string sound from a specified frequency. To simplify this process, we created a \verb|Note| class that encapsulates both the frequency and amplitude parameters. All active notes are then stored in a buffer that is accessed by separate threads to add notes, accumulate the final audio output, and remove notes once they have been played. The buffer ensures that data is available when required, allowing the visual and audio threads to operate independently without interference or data loss.

Our complete source code is available at: \url{https://github.com/Egglis/LineHarp}

\section{Results and Discussion}\label{sec:results}

\begin{figure}[tb]
 \centering 
 \includegraphics[width=\columnwidth]{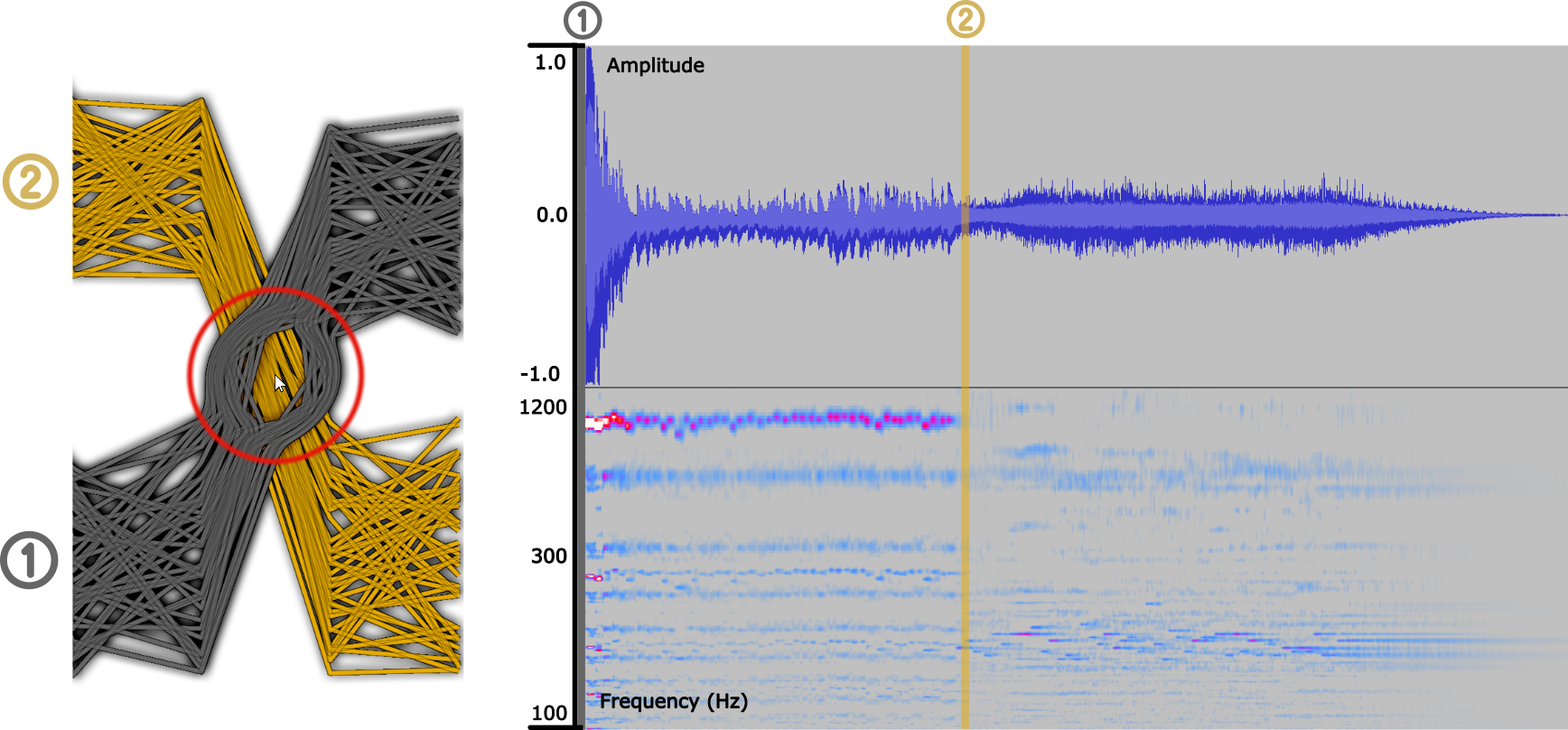}
 \caption{Left: visualization of two overlapping clusters, with the grey cluster visually displaced. Right: audio visualization of all lines intersecting the lens (red circle) played in order of importance, with a 0.05$\mskip\thinmuskip$s delay between each note (1200$\mskip\thinmuskip$bpm).}
 \label{fig:exampleLens}
\end{figure}

In order to visualize the audio produced by our method, we employ the popular software package Audacity to record and visualize the audio output. In addition to common waveform visualizations, which plot amplitude over time (as shown, for example, in Figures \ref{fig:comparison} and \ref{fig:combined}), we also utilize Audacity's spectrogram view using the Enhanced Autocorrelation (EAC) algorithm~\cite{audacity}, as shown in Figure~\ref{fig:angleMapping} as well as the bottom right of Figures~\ref{fig:teaser} and \ref{fig:exampleLens}. The spectrogram highlights the contour of the fundamental frequency (musical pitch) of the audio, characterizing the perceived frequency of the sound. Additionally, we refer to the supplementary material for an audiovisual demonstration of our approach.

Identifying clusters in dense line charts and parallel coordinates is a common task, and analyzing their trends and features is equally vital in many cases. We use an example dataset that contains four distinct clusters with varying features, as illustrated in Figure \ref{fig:teaser}. The dataset also includes 150 random lines (dark green), which are of lower importance compared to the clustered lines. The speed of the mouse movement is an important factor in determining the level of detail in our sonification approach. Faster mouse movement will produce a more generalized overview, while a slower movement will generate more detailed audio feedback. In Figure \ref{fig:teaser}, we used a fixed mouse movement speed for a duration of 5 seconds indicated by the red line. The audio output resulting from this mouse movement generates audio feedback from the clusters in the following order: purple, yellow, gray, and brown. The additional audio visualization in Figure \ref{fig:teaser} provides a visual representation of the audible feedback produce by this mouse movement. By examining the visualization, we can see that the amplitude spikes whenever the mouse transitions from one cluster to another, and that the corresponding frequency also changes according to our mapping. Based on the visualization of the audio and the perception of the actual sound, we argue that clusters are distinguishable based on their audio feedback. We also conclude that frequency can be used to help perceive angles and analyze trends and features.

When clusters overlap, the audio feedback generated by our sonification approach prioritizes the most important lines. However, this is not ideal when we want to analyze obscured clusters, as shown in Figure \ref{fig:exampleLens}, where the gray cluster is rendered on top of the yellow cluster. For the visual channel this can be solved with the use of lenses that displace overlapping clusters \cite{Toolglass, Distortion3D}. Similarly, the audio channel also requires a function to filter and focus the audio feedback. Figure \ref{fig:exampleLens} shows how our lens feature can be utilized for filtering and focusing, based on a section of a dataset that has two overlapping clusters.

To operate our lens, users can adjust the radius using the mouse and manipulate the importance threshold with key presses. When the threshold is set to visually reveal the yellow cluster, as shown on the left-hand side of Figure \ref{fig:exampleLens}, the corresponding sonification ignores the displaced lines. By gradually altering the threshold, users can inspect the lens region in more detail. Furthermore, using our lens playback feature, all line segments contained within the lens radius are played back in importance order. The resulting audio output, visualized on the right-hand side of Figure \ref{fig:exampleLens}, includes both amplitude and frequency. We see that the audio feedback changes frequency when the iteration reaches the lower second cluster (yellow), allowing users to distinguish between overlapping clusters and providing an indication for their overall directionality as well as their homogeneity.

Based on our experiments, we believe that our approach can support the perception of density variations and clusters in line charts. Furthermore, we believe that the addition of a sonified lens can further improve the sonification by filtering the audio feedback. However, to substantiate our claims, empirical evaluations and comparisons are required. Previous research already suggests that an interactive approach to sonification of line charts is preferable for complex data visualization over direct mappings \cite{TheSonificationHandbook}. For instance, a study that aimed to identify high density areas in complex line charts found that sonification can improve accuracy \cite{SONIFICATION_DENSEDATA}. 

Like many sonification models based on complex data visualizations, training is often required for optimal performance. Unlike visual charts, which most humans are taught how to read from a very young age through formal education systems, auditory charts require longer training for optimal usage. While our approach aims to provide an intuitive sonification model, it still suffers from the fact that the audio feedback can be difficult to interpret at times. However, studies have shown that training can improve accuracy in point estimation tasks and may even lead to better performance than a standard visual graph with sufficient training \cite{point_estimation_sonification_task, walkerbrief}. How training might impact the accuracy of our model requires further evaluation, but generally accuracy should improve over time.  

Finally, our approach suffers from a lack of auditory context (such as axes or tick marks) when using a mouse interface, which can make orientation difficult. This can be partially remedied by using a touch screen, where a user will be able detect boundaries and location due to human proprioception, the ability to sense self-movement, force, and body position \cite{Proprioception}. This has also been found to be just as effective for people with reduced vision \cite{Balance_in_the_Blind}. 

\section{Conclusion}

In this work, we introduced Line Harp, a new sonification approach for dense line charts that combines interactive audio feedback with visual representations of data. We presented an importance-driven sonification method that combines a frequency-based encoding of line direction with interactive lenses and provides a way to focus the audio output. Our directional frequency mapping supports line chart angle perception while also serving as an aid for chart navigation. Furthermore, we proposed a technique  that dynamically scales amplitudes to emphasize clustered lines and reduce the overall influence of less important lines to improve density perception. Overall, Line Harp is a promising tool that could potentially enhance the accessibility of visualizations for individuals with visual impairments, or provide a more immersive experience for data analysts. In future work, we aim to delve deeper into the realm of sonification for data analysis, including conducting empirical studies, in order to fully explore and harness its potential for enhancing accessibility and perception of dense visualizations.

\bibliographystyle{abbrv-doi}

\bibliography{template}

\begin{thebibliography}{10}

\bibitem{audacity}
Audacity.
\newblock Spectrogram view.
\newblock \url{https://manual.audacityteam.org/man/spectrogram_view.html}.
\newblock Accessed: March, 2023.

\bibitem{Toolglass}
E.~A. Bier, M.~C. Stone, K.~Pier, W.~Buxton, and T.~D. DeRose.
\newblock Toolglass and magic lenses: The see-through interface.
\newblock In {\em Proceedings of ACM SIGGRAPH}, pp. 73--80, 1993. doi: {{%
10\hspace{.1pt}\discretionary{.}{%
}{.}\hspace{.4pt}1145\discretionary{/}{%
}{/}166117\hspace{.1pt}\discretionary{.}{%
}{.}\hspace{.4pt}166126}}


\bibitem{TANGIBLESCANNING}
T.~Bovermann, T.~Hermann, and H.~J. Ritter.
\newblock Tangible data scanning sonification model.
\newblock In {\em Proceedings of ICAD}, pp. 77--82, 2006.

\bibitem{guidelinesForEarcons}
S.~A. Brewster, P.~C. Wright, and A.~D. Edwards.
\newblock Experimentally derived guidelines for the creation of earcons.
\newblock In {\em Proceedings of HCI}, pp. 155--159, 1995.

\bibitem{BrowsingModes}
L.~Brown, S.~Brewster, R.~Ramloll, W.~Yu, and B.~Riedel.
\newblock Browsing modes for exploring sonified line graphs.
\newblock In {\em Proceedings of BCS HCI}, vol.~2, pp. 6--9, 2002.

\bibitem{Drawing_by_Ear}
L.~M. Brown and S.~A. Brewster.
\newblock Drawing by ear: Interpreting sonified line graphs.
\newblock In {\em Proceedings of ICAD}, pp. 152--156, 2003.

\bibitem{Design_Guidelines}
L.~M. Brown, S.~A. Brewster, S.~Ramloll, R.~Burton, and B.~Riedel.
\newblock Design guidelines for audio presentation of graphs and tables.
\newblock In {\em Proceedings of ICAD}, pp. 284--287, 2003.

\bibitem{Bruckner-2010-HVC}
S.~Bruckner, P.~Rautek, I.~Viola, M.~Roberts, M.~C. Sousa, and M.~E.
  Gr{\"o}ller.
\newblock Hybrid visibility compositing and masking for illustrative rendering.
\newblock {\em Computers \& Graphics}, 34(4):361--369, 2010. doi: {{%
10\hspace{.1pt}\discretionary{.}{%
}{.}\hspace{.4pt}1016\discretionary{/}{%
}{/}j\hspace{.1pt}\discretionary{.}{%
}{.}\hspace{.4pt}cag\hspace{.1pt}\discretionary{.}{%
}{.}\hspace{.4pt}2010\hspace{.1pt}\discretionary{.}{%
}{.}\hspace{.4pt}04\hspace{.1pt}\discretionary{.}{%
}{.}\hspace{.4pt}003}}


\bibitem{AirTraffic}
D.~Cabrera, S.~Ferguson, and G.~W. Laing.
\newblock Considerations arising from the development of auditory alerts for
  air traffic control consoles.
\newblock In {\em Proceedings of ICAD}, pp. 242--245, 2005.

\bibitem{Distortion3D}
M.~S.~T. Carpendale, D.~J. Cowperthwaite, and F.~D. Fracchia.
\newblock Distortion viewing techniques for 3-dimensional data.
\newblock In {\em Proceedings of IEEE InfoVis}, pp. 46--53, 1996. doi: {{%
10\hspace{.1pt}\discretionary{.}{%
}{.}\hspace{.4pt}1109\discretionary{/}{%
}{/}INFVIS\hspace{.1pt}\discretionary{.}{%
}{.}\hspace{.4pt}1996\hspace{.1pt}\discretionary{.}{%
}{.}\hspace{.4pt}559215}}


\bibitem{Balance_in_the_Blind}
H.~Daneshmandi, A.~Norasteh, and H.~Zarei.
\newblock Balance in the blind: A systematic review.
\newblock {\em Physical Treatments: Specific Physical Therapy Journal},
  11:1--12, 2021. doi: {{%
10\hspace{.1pt}\discretionary{.}{%
}{.}\hspace{.4pt}32598\discretionary{/}{%
}{/}ptj\hspace{.1pt}\discretionary{.}{%
}{.}\hspace{.4pt}11\hspace{.1pt}\discretionary{.}{%
}{.}\hspace{.4pt}1\hspace{.1pt}\discretionary{.}{%
}{.}\hspace{.4pt}430\hspace{.1pt}\discretionary{.}{%
}{.}\hspace{.4pt}2}}


\bibitem{SoniScope20221095}
K.~Enge, A.~Rind, M.~Iber, R.~Höldrich, and W.~Aigner.
\newblock Towards multimodal exploratory data analysis: Soniscope as a
  prototypical implementation.
\newblock In {\em Proceedings of EuroVis (Short Papers)}, pp. 67--71, 2022.
  doi: {{%
10\hspace{.1pt}\discretionary{.}{%
}{.}\hspace{.4pt}2312\discretionary{/}{%
}{/}evs\hspace{.1pt}\discretionary{.}{%
}{.}\hspace{.4pt}20221095}}


\bibitem{THIRTEENYO}
J.~H. Flowers.
\newblock Thirteen years of reflection on auditory graphing: Promises,
  pitfalls, and potential new directions.
\newblock In {\em Proceedings of ICAD}, pp. 406--409, 2005.

\bibitem{VisualGraphs}
J.~H. Flowers and T.~A. Hauer.
\newblock Musical versus visual graphs: Cross-modal equivalence in perception
  of time series data.
\newblock {\em Human Factors: The Journal of Human Factors and Ergonomics
  Society}, 37(3):553--569, 1995. doi: {{%
10\hspace{.1pt}\discretionary{.}{%
}{.}\hspace{.4pt}1518\discretionary{/}{%
}{/}001872095779049264}}


\bibitem{STAR_PC}
J.~Heinrich and D.~Weiskopf.
\newblock State of the art of parallel coordinates.
\newblock In {\em Proceedings of Eurographics (State of the Art Reports)}, pp.
  95--116, 2013. doi: {{%
10\hspace{.1pt}\discretionary{.}{%
}{.}\hspace{.4pt}2312\discretionary{/}{%
}{/}conf\discretionary{/}{%
}{/}EG2013\discretionary{/}{%
}{/}stars\discretionary{/}{%
}{/}095\discretionary{%
}{-}{-}116}}


\bibitem{EEG}
T.~Hermann, G.~Baier, U.~Stephani, and H.~Ritter.
\newblock Vocal sonification of pathologic eeg features.
\newblock In {\em Proceedings of ICAD}, pp. 158--163, 2006.

\bibitem{TheSonificationHandbook}
T.~Hermann, A.~Hunt, and J.~G. Neuhoff.
\newblock {\em The Sonification Handbook}.
\newblock Logos Verlag Berlin, 1 ed., 2011.

\bibitem{Hermann2000PrincipalCS}
T.~Hermann, P.~Meinicke, and H.~J. Ritter.
\newblock Principal curve sonification.
\newblock In {\em Proceedings of ICAD}, pp. 81--86, 2000.

\bibitem{Model-Based}
T.~Hermann and H.~Ritter.
\newblock Listen to your data: Model-based sonification for data analysis.
\newblock In {\em Advances in Intelligent Computing and Multimedia Systems},
  pp. 189--194. International Institute for Advance Studies in Systems Research
  and Cybernetics, 1999.

\bibitem{Marketbuzz}
P.~Janata and E.~Childs.
\newblock Marketbuzz: Sonification of real-time financial data.
\newblock In {\em Proceedings of ICAD}, 2004.

\bibitem{ClockingTheMind}
A.~R. Jensen.
\newblock Reaction time as a function of experimental conditions.
\newblock In {\em Clocking the Mind}, pp. 43--54. Elsevier Science Ltd, 2006.
  doi: {{%
10\hspace{.1pt}\discretionary{.}{%
}{.}\hspace{.4pt}1016\discretionary{/}{%
}{/}B978\discretionary{%
}{-}{-}008044939\discretionary{%
}{-}{-}5\discretionary{/}{%
}{/}50004\discretionary{%
}{-}{-}5}}


\bibitem{AccessibilityWild}
S.~C.~S. Joyner, A.~Riegelhuth, K.~Garrity, Y.-S. Kim, and N.~W. Kim.
\newblock Visualization accessibility in the wild: Challenges faced by
  visualization designers.
\newblock In {\em Proceedings of ACM CHI}, pp. 1--19, 2022. doi: {{%
10\hspace{.1pt}\discretionary{.}{%
}{.}\hspace{.4pt}1145\discretionary{/}{%
}{/}3491102\hspace{.1pt}\discretionary{.}{%
}{.}\hspace{.4pt}3517630}}


\bibitem{Music_Perception_2002}
T.~Justus and J.~Bharucha.
\newblock Music perception and cognition.
\newblock In {\em Stevens' Handbook of Experimental Psychology}, pp. 453--492.
  John Wiley \& Sons Inc, 2002. doi: {{%
10\hspace{.1pt}\discretionary{.}{%
}{.}\hspace{.4pt}1002\discretionary{/}{%
}{/}0471214426\hspace{.1pt}\discretionary{.}{%
}{.}\hspace{.4pt}pas0111}}


\bibitem{AccessibleVisualization}
N.~W. Kim, S.~C. Joyner, A.~Riegelhuth, and Y.~Kim.
\newblock Accessible visualization: Design space, opportunities, and
  challenges.
\newblock {\em Computer Graphics Forum}, 40(3):173--188, 2021. doi: {{%
10\hspace{.1pt}\discretionary{.}{%
}{.}\hspace{.4pt}1111\discretionary{/}{%
}{/}cgf\hspace{.1pt}\discretionary{.}{%
}{.}\hspace{.4pt}14298}}


\bibitem{Ramloll2001UsingNS}
R.~Ramloll, S.~A. Brewster, W.~Yu, and B.~Riedel.
\newblock Using non-speech sounds to improve access to {2D} tabular numerical
  information for visually impaired users.
\newblock In {\em Proceedings of BCS HCI/IHM}, pp. 515--529, 2001. doi: {{%
10\hspace{.1pt}\discretionary{.}{%
}{.}\hspace{.4pt}1007\discretionary{/}{%
}{/}978\discretionary{%
}{-}{-}1\discretionary{%
}{-}{-}4471\discretionary{%
}{-}{-}0353\discretionary{%
}{-}{-}0\_32}}


\bibitem{SONIFICATION_DENSEDATA}
N.~Rönnberg and J.~Johansson~Westberg.
\newblock Interactive sonification for visual dense data displays.
\newblock In {\em Proceedings of ISon}, pp. 63--67, 2016.

\bibitem{medical}
P.~Sanderson, S.~Eunice, L.~Philippe, and W.~Alexandra.
\newblock Auditory alarms, medical standards, and urgency.
\newblock In {\em Proceedings of ICAD}, pp. 24--27, 2006.

\bibitem{point_estimation_sonification_task}
D.~R. Smith and B.~N. Walker.
\newblock Effects of auditory context cues and training on performance of a
  point estimation sonification task.
\newblock {\em Applied Cognitive Psychology}, 19(8):1065--1087, 2005. doi: {{%
10\hspace{.1pt}\discretionary{.}{%
}{.}\hspace{.4pt}1002\discretionary{/}{%
}{/}acp\hspace{.1pt}\discretionary{.}{%
}{.}\hspace{.4pt}1146}}


\bibitem{Interactive_Lenses}
C.~Tominski, S.~Gladisch, U.~Kister, R.~Dachselt, and H.~Schumann.
\newblock Interactive lenses for visualization: An extended survey.
\newblock {\em Computer Graphics Forum}, 36(6):173--200, 2017. doi: {{%
10\hspace{.1pt}\discretionary{.}{%
}{.}\hspace{.4pt}1111\discretionary{/}{%
}{/}cgf\hspace{.1pt}\discretionary{.}{%
}{.}\hspace{.4pt}12871}}


\bibitem{Tominski20IVDA}
C.~Tominski and H.~Schumann.
\newblock {\em Interactive Visual Data Analysis}.
\newblock AK Peters Visualization Series. CRC Press, 2020. doi: {{%
10\hspace{.1pt}\discretionary{.}{%
}{.}\hspace{.4pt}1201\discretionary{/}{%
}{/}9781315152707}}


\bibitem{LineWeaver}
T.~Trautner and S.~Bruckner.
\newblock Line weaver: Importance-driven order enhanced rendering of dense line
  charts.
\newblock {\em Computer Graphics Forum}, 40(3):399--410, 2021. doi: {{%
10\hspace{.1pt}\discretionary{.}{%
}{.}\hspace{.4pt}1111\discretionary{/}{%
}{/}cgf\hspace{.1pt}\discretionary{.}{%
}{.}\hspace{.4pt}14316}}


\bibitem{Tnnermann2009MultitouchIF}
R.~T{\"u}nnermann, Kolbe, T.~L. Bovermann, and T.~Hermann.
\newblock Surface interactions for interactive sonification.
\newblock In {\em Proceedings of ICAD}, pp. 166--–183, 2009. doi: {{%
10\hspace{.1pt}\discretionary{.}{%
}{.}\hspace{.4pt}1007\discretionary{/}{%
}{/}978\discretionary{%
}{-}{-}3\discretionary{%
}{-}{-}642\discretionary{%
}{-}{-}12439\discretionary{%
}{-}{-}6\_9}}


\bibitem{Proprioception}
J.~C. Tuthill and E.~Azim.
\newblock Proprioception.
\newblock {\em Current Biology}, 28(5):194--203, 2018. doi: {{%
10\hspace{.1pt}\discretionary{.}{%
}{.}\hspace{.4pt}1016\discretionary{/}{%
}{/}j\hspace{.1pt}\discretionary{.}{%
}{.}\hspace{.4pt}cub\hspace{.1pt}\discretionary{.}{%
}{.}\hspace{.4pt}2018\hspace{.1pt}\discretionary{.}{%
}{.}\hspace{.4pt}01\hspace{.1pt}\discretionary{.}{%
}{.}\hspace{.4pt}064}}


\bibitem{SonificationMappings2}
B.~Walker.
\newblock Magnitude estimation of conceptual data dimensions for use in
  sonification.
\newblock {\em Journal of Experimental Psychology: Applied}, 8(1):211--21,
  2003. doi: {{%
10\hspace{.1pt}\discretionary{.}{%
}{.}\hspace{.4pt}1037\discretionary{/}{%
}{/}1076\discretionary{%
}{-}{-}898X\hspace{.1pt}\discretionary{.}{%
}{.}\hspace{.4pt}8\hspace{.1pt}\discretionary{.}{%
}{.}\hspace{.4pt}4\hspace{.1pt}\discretionary{.}{%
}{.}\hspace{.4pt}211}}


\bibitem{SonificationMappings}
B.~Walker and D.~Lane.
\newblock Psychophysical scaling of sonification mappings: A comparison of
  visually impaired and sighted listeners.
\newblock In {\em Proceedings of ICAD}, pp. 90--94, 2001.

\bibitem{walkerbrief}
B.~N. Walker and M.~A. Nees.
\newblock Brief training for performance of a point estimation sonification
  task.
\newblock In {\em Proceedings of ICAD}, pp. 276--279, 2005.

\bibitem{TheoryOfSonification}
B.~N. Walker and M.~A. Nees.
\newblock Theory of sonification.
\newblock In {\em The Sonification Handbook}, pp. 9--39. Logos Verlag Berlin, 1
  ed., 2011.

\bibitem{EdgeLens}
N.~Wong, S.~Carpendale, and S.~Greenberg.
\newblock Edgelens: an interactive method for managing edge congestion in
  graphs.
\newblock In {\em Proceedings of IEEE InfoVis}, pp. 51--58, 2003. doi: {{%
10\hspace{.1pt}\discretionary{.}{%
}{.}\hspace{.4pt}1109\discretionary{/}{%
}{/}INFVIS\hspace{.1pt}\discretionary{.}{%
}{.}\hspace{.4pt}2003\hspace{.1pt}\discretionary{.}{%
}{.}\hspace{.4pt}1249008}}


\end{thebibliography}
\end{document}